# A simple approach to the correlation of rotovibrational states in four-atomic molecules

**N.Manini**[1,2], **S.Oss**[2]

[1]Scuola Internazionale Superiore di Studi Avanzati, 34014 Trieste, ̋Italy

[2]Dipartimento di Fisica, Università di Trento and INFM, 38050 ̋Povo (TN), Italy

(S.Oss, phone +39-461-881633, FAX +39-461-881696, e.mail ̋oss@itnsg1.science.unitn.it)

**Abstract**

The problem of correlation between quantum states of four-atomic molecules in different geometrical configurations is reviewed in detail. A general, still simple rule is obtained which allows one to correlate states of a linear four-atomic molecule with those of any kind of non-linear four-atomic molecule.

**PACS**: 33.10.Cs, 33.10.Jz

## 1. Introduction

In the past few years a considerable attention has been devoted to the spectroscopical analysis of several four-atomic molecules. From the experimental point of view, new high resolution spectra (both Raman and infrared) have been obtained [1-5] showing very complex roto-vibrational patterns not yet completely explained. On the theoretical side, correspondingly, a lot of work is in progress: not so surprisingly, going from three- to four-atomic molecules leads very often to difficult situations in which it can be simply impossible to describe adequately the molecular dynamics. Molecules can in fact have distinct equilibrium shapes, depending on the form of their potential energy function. Such potential function can be ideally transformed from one form into another. It is then important to observe how the rotovibrational states correspondingly change in this transformation. This is an essentially complicated problem, because of the different equilibrium shapes available to a four-atomic molecule. Moreover, the way of correlating states is not necessarily unique, as it may depend on the path used to carry out the potential surface transformation. As a consequence, further degrees of freedom (like vibrational torsional modes [6,7]) can become accessible to the molecule. In the three-atomic case, a simple way is available since a very long time [8] to correlate vibrational states in the linear configuration with states characterizing the bent molecular geometry. In the four-atomic case the same problem has been studied and solved in the framework of group-theoretical methods [9,10] just in the special case of symmetric four-atomic molecules. A general rule allowing one to construct a complete correlation diagram for *any* kind of four-atomic molecule is not yet available. The idea of realizing a correlation between rotovibrational states of molecules in different equilibrium shapes is of extreme importance for example in the construction of simple parametrized forms of Hamiltonian operators. In this paper we extend the known results holding for symmetrical four-atomic molecules to the non-symmetrical case. This work will not make use of group-theoretical methods, which are in fact basically useless when dealing with lowered molecular symmetries. With our simple method, moreover, some explicit formulas for the correlation laws between rotovibrational quantum numbers are obtained for basically any relevant molecular shapes and symmetries. This is another important result, as it allows one to study the correlation problem from a more systematic point of view.

The present paper is structured as follows: after a section in which the description of the simple correlation laws for the three-atomic case is briefly summarized, the problem for four-atomic molecules is attacked in a distinct section. Here the specific aspects of free and hindered internal motions are considered.

**2. Correlation for three-atomic molecules**

The idea of correlation of rotovibrational states between different molecular shapes finds its basis on the fact that any physical system can be described by means of some parametrized form of a suitable Hamiltonian operator. The dynamics of the physical system is thus dependent on the values assumed by a set of continuously varying parameters. The effect of small variations in these parameters can usually be treated by some perturbative technique. The specific case of molecular correlation is however a non-perturbative problem, in which the continuous change of one (or more) parameter over its whole range gives rise to substantial modifications in the potential surface and, consequently, in the dynamics of at least one molecular degree of freedom. It must be stressed that such non-perturbative effects cannot be continuously controlled in an experiment [9]. On consequence, correlation models are difficult to be realized and tested.

Consider for example the potential function of a linear molecule. Let $\alpha$ be the bending angle giving the deviation from the linear equilibrium position of the molecule. The potential will be some anharmonic function of $\alpha$, symmetric with respect to the minimum position reached at $\alpha=0$. A completely different situation occurs when the molecule, in its equilibrium configuration, is bent. In this case a potential barrier is expected with its maximum value at $\alpha=0$. These two cases, linear and bent molecule, can be continuously connected (i.e. their eigenstates and eigenvalues correlate) by means of a potential barrier whose height is varying from zero to a given maximum value. This correspondence is realized in practice with an Hamiltonian operator containing a parameter related to the height of the barrier. Intermediate or small values of this parameter will describe quasi-linear molecules, in which the equilibrium shape is only slightly bent. In such molecules the vibrational mode in the bending angle still recalls that of a linear molecule.

The correlation scheme of a triatomic molecule, as discussed in [8], is easily obtained after considering the rotovibrational states and their quantum labels. The six rotovibrational degrees of freedom of a linear three-atomic molecule are labeled by $J$, $M$ (for the rotational part), $v_1$ and $v_2$ (for stretching vibrations) and $v_b$, $l$ (for bending vibrations). A non-linear top can be described by $J$, $M$, $k$ (for the rotational part), $v_1$ and $v_2$ (for stretching vibrations) and $v_b$ (for bending vibrations). A remarkable feature is the conversion of a vibrational quantum number for the linear case, $l$, in a rotational quantum number for the bent case, $k$. This is easily explained in the symmetric top approximation by observing that the allowed values˝of the total angular momentum quantum number $J$ are the following:

$$J=|l|, |l|+1, |l|+2, \ldots \text{ (linear top)}; \quad J=|k|, |k|+1, |k|+2, \ldots \text{ (symmetric top)}. \qquad (1)$$

As a consequence, the first formula for the correlation scheme will be simply given by $k=l$. Stretching vibrations correlate in a simple way, as $v_1$ and $v_2$ quantum numbers result to label corresponding states for both linear and bent molecules. The same is also true for the rotational labels $J$ and $M$. A less obvious correlation formula is established for the bending vibrational mode. As it is shown in Fig.1, going from a linear to a bent configuration, those states (labeled by $v_b^{|l|}$) with $v_b=|l|$ move to construct the rotational band of the ground state ($v_b=0$) for the bent molecule. A general relation for the correlation of bending states can be˝expressed as

$$v_b(\text{bent molecule}) = \frac{v_b(\text{linear molecule}) - |l|}{2}. \qquad (2)$$

This simple correlation scheme is very important. It forms, in fact, the starting point for the extension to the four-atomic case, as it will be discussed in˝the following section.

## 3. Correlation for four-atomic molecules

The problem of the correlation of rotovibrational states for four-atomic molecules is by no means a trivial one [9,10]. It happens in fact that, starting from a linear configuration, several non-equivalent molecular geometries arise. The dynamics of these different physical systems correlate among themselves following complex rules. In order to explore in detail all such possibilities, we find it useful to divide non-linear four-atomic molecules in two families. In the first one chain molecules are considered, in which each atom is involved in no more that two chemical bonds. The second family includes those molecules with one central atom sharing its chemical bonds with three atoms. We shall consider this kind of molecules later.

Chain molecules can be further divided, according to their geometry, in the following groups (see Fig.2): (*i*) linear molecules (for example HCCH, HCCF); (*ii*) half-linear molecules: the equilibrium angle is less then 180 degrees for only one of the two bending coordinates (HCNO); (*iii*) bent molecules with free internal rotation: this kind of molecules is obtained by considering both bending angles less than 180 degrees at equilibrium. As it will be discussed later in greater detail, it is convenient to consider the ideal case in which the rotation of one molecular section is free with respect to the other section; (*iv*) planar bent molecules: internal rotation is now affected by some kind of potential energy (hindered rotation). The minimum energy is achieved in correspondence with a planar geometry (HNCO); (*v*) non-planar bent molecules: like in the planar case, the internal rotation suffers the effects of a potential energy. The minimum energy is now reached in a non-planar configuration, thus allowing for two equivalent non-superimposable equilibrium geometries (HOOH).

Chain molecules can be classified also according to a different criterion, based on the symmetry with respect to bond exchange. More specifically, symmetrical molecules of the type ACCA are thoroughly studied in [10]. We will now consider the non-symmetrical case of four-atomic molecules like ACDB from a completely different point of view, that is, not making use of group-theoretical statements. In the following analysis we will focus our attention on the most important properties of only those degrees of freedom relevant in the description of the problem of interest.

We address this problem by starting from the special case in which the two central atoms in the molecular chain have infinitely large masses and intramode couplings are all negligible. In this way our arguments take advantage of a local or quasi-local molecular picture and only those problems related to the specific aspects of the correlation scheme result of interest here. Moreover, axis-switching phenomena and rotovibrational couplings have negligible contributions to the overall dynamics of the system and the axis defined by the heavy atoms remains in any geometrical configuration a good inertial principal axis. The more realistic case in which all atoms have finite masses can be easily obtained by means of simple considerations we defer to section 3.3.

As a first step, it is necessary to observe that the 9 rotovibrational degrees of freedom of a four-atomic molecule are divided in 2 rotations and 7 vibrations for a linear molecule and 3 rotations and 6 vibrations for a non-linear molecule. Like in the case of three-atomic molecules, rotational quantum numbers $J$ and $M$ correlate with the corresponding quantities going from linear to non-linear molecules. The same happens for the 3 stretching vibrations. For this reason we need to focus our attention only on the remaining degrees of freedom, which are related to bending vibrations.

The correlation for these four bending degrees of freedom will be now established by relating it to the simpler three-atomic case. The key point is that any chain four-atomic molecule can be thought as constructed by two three-atomic molecules each of them "sharing" one of its two bonds, as shown in Fig.3. As a result, if one neglects the local modes coupling, the two benders of the four-atomic molecule act in the same way as the two separate three-atomic benders. If one labels quantum states of the two benders as $|v_4^{l_4} v_5^{l_5}>$ (in the linear configuration), the correlation law given in Eq. (2) can be readily extended˝to the more general expression

$$v_A = \frac{v_4 - |l_4|}{2}, \quad k_A = l_4 ,$$
$$v_B = \frac{v_5 - |l_5|}{2}, \quad k_B = l_5 , \qquad (3)$$

in which $v_A$, $v_B$ denote vibrational quantum numbers for local bending motions of the non-linear molecule and $k_A$, $k_B$ label the projections of the angular momentum of both benders along the axis defined by the two heavy nuclei. We are using here arbitrary labels for bent geometry to stress the purely local meaning of the corresponding degrees of freedom. Labels $v_A$, $v_B$, for example, quantize the uncoupled bending modes of the two triatomic˝molecular sections. These purely local modes have of course to be thought as an ideal limiting situation used here to clarify some basic concepts.

In order to properly describe the meaning of the correlation law (3), it is necessary to apply it separately to different molecular equilibrium geometries.

*3.1. Half-linear molecules*

Several cases of half-linear molecules have been studied elsewhere [9]. Since one of the two triatomic sections is linear at equilibrium, no internal rotation of one section relative to the other can take place. Due to the small deviation from linearity of the bent section, some of those half-linear systems can be rather classified as quasi-linear molecules. For such molecules, a convenient labeling scheme is that of a linear molecule, $|v_4^{l_4} v_5^{l_5}>$. In case of strong deviation from linearity, it may be more convenient to partly apply correlation (3), by introducing an hybrid labeling scheme $|v_4^{l_4} v_b, k>$, in which $v_b=(v_5-|l_5|)/2$ and $k=l=l_4+l_5$. In ref. [9] a parameter of quasi-linearity is defined which allows to establish in a quantitative fashion the more appropriate labels for each specific case. The complete labeling should include, of course, three labels for stretching vibrations and the total angular momentum $J$. These quantum numbers maintain their original meaning in both linear and bent molecules.

*3.2. Bent molecules: free internal rotations*

As already suggested at the beginning of this section, in a bent four-atomic molecule an internal torsional mode can be excited. This mode arises naturally when considering the physical meaning of the $k_A$ and $k_B$ quantum numbers of Eq. (3) for a molecule with both triatomic sections with a bent geometry. These quantum numbers label the projections of the angular momenta of the light atoms along the axis of the heavy atoms. The rotations of the light atoms about this axis can be thought initially as free: no privileged direction appears moving away from linearity. As a further step, this torsional mode will be hindered by introducing the effect of a potential barrier. The natural coordinate for such degree of freedom is the dihedral angle $\chi$ as defined in Fig.4. A complete understanding of the dynamics of hindered rotations can be achieved after considering in some detail the proper quantum mechanical treatment of free rotations (see also [7] and [11]).

The two triatomic sections of the molecule can be seen as two different symmetric tops A, B free to rotate around the same axis fixed in space. If these tops are not interacting, a possible wavefunction of this system can be given as in

$$\psi_{k_A,k_B}(\theta_A,\theta_B) = C\, e^{i(k_A\theta_A + k_B\theta_B)}, \qquad (4)$$

where $\theta_A$, $\theta_B$ are the angles defined in Fig.4 and $k_A$, $k_B$ are labels for the projection of angular momentum operators along the common axis of rotation. These quantum numbers are naturally identified with those defined in Eq. (3). A more convenient factorization of the rotational wavefunction is obtained by considering the following linear transformation:

$$\chi = \theta_A - \theta_B, \quad \Omega = \frac{\theta_A + \theta_B}{2}. \qquad (5)$$

In terms of the angles $\chi$, $\Omega$ the wavefunction (4) can be written as

$$\psi_{k,k_i}(\chi,\Omega) = C\, e^{i\left(k\Omega + k_i \frac{\chi}{2}\right)}, \qquad (6)$$

where the quantum numbers $k$, $k_i$ are defined by

$k = k_A + k_B, \quad k_i = k_A - k_B.$  (7)

The wavefunction (6) can be thought as factored in two parts. The first one, depending on $\Omega$, contributes to the overall rotational wavefunction. Its quantum number $k$ is the usual label for the projection of the total molecular angular momentum operator. The second part, depending on $\chi$, describes the internal (free) rotation of the molecule. The definition (7) of $k$, $k_i$ has as an obvious consequence the fact that these labels have the same parity. This implies that rotational bands starting from a given $k_i$ state are composed either by even-$k$ states (for $k_i$ even) or by odd-$k$ states (for $k_i$ odd). In particular, only states of even internal rotation will contribute to the pure vibrational spectrum of the bent molecule.

Another point worth mentioning here is that, according to Eqs. (5) and (6), the variable $\chi$ has an angular periodicity of $4\pi$. This can be clearly seen in Fig.5, where a possible configuration of a planar bent four-atomic molecule is shown for different values of $\chi$, $\Omega$ (or $\theta_A$ and $\theta_B$). A change in the dihedral angle $\chi$ of $2\pi$ gives the original cis-configuration (Fig.5a) rigidly rotated through an angle of $\pi$ around the molecular axis (Fig.5c), although $\Omega$ remains unchanged. A further change of $\chi$ through an angle of $2\pi$ is needed in order to recover the original configuration inclusive of the overall rotation (Fig.5e, but see also Fig.1˝in [7]).

These arguments turn out to be of indispensable help in the following discussion concerning hindered internal rotations.

*3.3. Bent molecules: hindered rotations*

The aim of this paragraph is to combine the main correlation law (3) (leading to an ideal scheme of free internal rotations) with that of free to hindered internal rotations, to obtain a complete correlation for "rigid" bent molecules. Such scheme must be capable to describe rotovibrational patterns of basically any kind of bent four-atomic molecule. The connection between free and hindered rotations has been well known for a long time [12]. Here we limit ourselves to a brief overview.

Consider a "rigid" planar bent molecule. Its stiffness is referred to the internal rotation which is now definitely hindered by a potential energy characterized by a deep minimum in correspondence of a planar (cis or trans) configuration. The correlation with the corresponding free-rotating molecule is obtained by considering the correct periodicity of the hindering potential in its dependence on the dihedral coordinate $\chi$, (see also Fig.5) which is given by

$$V(\chi)=V(\chi+2\pi). \tag{8}$$

This periodicity implies that the hindering potential function has at least two equivalent minima in the [0-4$\pi$] domain of the variable $\chi$. Such minima are readily recognized in Fig.5 as those given, in the cis-minimum case, by the two equivalent molecular configurations (a) and (c) differing only by an overall rotation through an angle of $\pi$.

In Fig.6 we report the continuous correlation scheme connecting the two extreme cases of a bent molecule with free internal rotation and rigid planar molecule. It is important to observe in this figure the origin of rotational bands superimposed to torsional levels of the rigid molecule. It is possible to establish a simple correlation formula for these torsional levels. Denoting by $v_T$ the quantum number of the torsional vibration, one has

$$v_T = k_i - \frac{(1-p)}{2}, \tag{9}$$

where $p=\pm 1$, depending on the parity ($\pm$) of the rotational sublevel of the free internal mode. It should be emphasized that the correlation scheme reproduced in Fig.6 can be used for any intermediate strength of the hindering potential. In cases like these, the internal rotation is only slightly hindered. As a consequence, tunneling amplitude through the potential barrier is still relevant. For very highly excited torsional states a more appropriate labeling could be that typical of free rotations, i.e. $|k,k_i\rangle$ in place of $|v_T,k\rangle$. It is important to notice that the differences between Fig.6 and Fig.165c in [12] are due to the equivalence of the light atoms in the case considered there ($C_2H_4$), as explained in [13].

The last case to be considered is that of bent non-planar molecules. For these molecules (such as DSSH, DOOH) two equilibrium configurations can be given, which cannot be superimposed by means of pure rotations but only by using inversion operations. For this reason the hindering potential surface will be characterized by two equivalent minima thus giving four minima in the $[0,4\pi]$ domain of the dihedral angle $\chi$. It is then possible to obtain the correlation scheme between planar and non-planar bent geometries by using this modified hindering potential function. Another, equivalent approach can be found in [13], based on the simple idea to raise a potential barrier in correspondence with the planar geometry. The resulting spectrum, depending on the relative height of the potential barrier, shows as a ″general feature inversion doublets related to the tunneling through the cis-trans barrier.

It is finally possible to use the general correlation formulas (3) and (9) to draw a global correlation scheme between rotovibrational states of a linear four-atomic molecule and those of a rigid bent molecule (either planar or non-planar). A basic version of such scheme is shown in Fig.7, in which only those states most relevant to our work (i.e. bending states of the linear molecule correlating to torsional states of the bent molecule)″are considered for clearness.

The initial restriction to chain molecules with two heavy central atoms can now be released. Such assumption is convenient in order to deal with local (mechanically uncoupled) modes. If atoms are all of comparable masses, local modes mixing must be accounted for. By continuously reducing the masses of the central atoms, vibrational wavefunctions will correspondingly move from a local to normal mode description: interactions among modes will make a pure local labeling scheme less and less meaningful. The latter is only a matter of notation, however, not affecting the physical mechanism driving the rotovibrational correlation pattern. As it is in fact well known, local modes interactions do not affect significantly those features of wavefunctions (e.g. the number of nodal points) relevant to the correlation of states. As a consequence, the results obtained for local modes still hold, with the only difference that correlating modes will result affected by other modes (in principle any stretching and/or bending mode can now interact, for example, with the torsional mode). Practical cases will depend on the specific molecule under examination, but the general correlation scheme shown in Fig.7 will maintain its qualitative validity, except that now labels can refer to normal modes.

The second generalization deals with the case of non-chain molecules, like ammonia and formaldehyde. It is again possible to apply the correlation formulas (3) and (9) used for chain molecules. Again, one has to imagine that a non-chain four-atomic molecule can be obtained by superimposing two bent three-atomic molecules, with the identification of one bond as shown in Fig.8. Consider for example the planar molecule of monodeuterated formaldehyde, HDCO. Its stretching modes come directly from the corresponding stretching modes of a fictitious linear HDCO molecule. Both in-plane bending motions and the torsional vibration related to the DCO and HCO planes are obtained from bending motions of DCO and HCO triatomic molecular sections, while the out-of-plane vibrational mode correlates with the torsional motion. A completely analogous approach is followed for non-planar molecules (like $NH_3$). These can be treated in the same way of chain non-planar molecules, involving pairs of equivalent minima of the potential function (thus explaining doubling of levels). Despite its apparent artificial construction, this scheme reveals itself of great help to include in the description of the rovibrational dynamics any kind of four-atomic molecule within the framework of the *same* parametrized Hamiltonian model.

*3.4. Symmetrical molecules*

As already pointed out, the case of symmetrical four-atomic molecules of the type ACCA is fully described in [10]. In this last paragraph we want to address the specific problem of the comparison between the correlation scheme as obtained in this previous work and the present results as given by equations (3) and (9) and shown in ″Fig.7.

The key feature related to symmetry is the degeneracy of the two local bending modes. Stretching modes are degenerate too, but their correlation (and that of *J* and *M* quantum numbers) is again a trivial problem not addressed here. This degeneracy leads to a complete mixing of wave functions for arbitrarily small interactions between local modes. The resulting coupled states will carry symmetrical and antisymmetrical representations of the permutational symmetry group labeled by *s* and *a* in [10]. We will now make an explicit correspondence between these symmetrized states and those states in a non-symmetrical molecule which are of interest here, that is, bending modes. We start by focusing our attention on the linear geometry. As pointed out in [10], the following bending Σ states must be considered in the construction of the correlation scheme:

$$\left(0^0 2^0\right)^+, \ \left(2^0 0^0\right)^+, \ \left(1^1 1^{-1}\right)^+, \ \left(1^1 1^{-1}\right)^-. \tag{10}$$

A further label u/g must be attached to these states in the symmetrical case, in which, moreover, a *normal* meaning is given to the traditional Herzberg labels $v_4^{l_4}$, $v_5^{l_5}$. The following four states are considered in the symmetric case:

$$\left(0^0 2^0\right)^+_g, \ \left(2^0 0^0\right)^+_g, \ \left(1^1 1^{-1}\right)^+_u, \ \left(1^1 1^{-1}\right)^-_u. \tag{11}$$

The translation between the local modes (10) and the normal modes (11) is by no means an obvious one due to a subtle notation problem. It can be easily shown in fact that, when approaching the normal limit, the highest $\Sigma_g^+$ state, say $(2^0 0^0)^+_g$, is characterized by a leading $(1^1 1^{-1})^+$ local component. The explanation of this effect is found in the level crossing due to the anharmonic splitting between $(1^1 1^{-1})^+$ and the pair $(2^0 0^0)^+, (0^0 2^0)^+$. The situation for the four states (11) can be summarized as follows:

| *symmetric normal notation* | *local notation* |
|---|---|
| $\left(0^0 2^0\right)^+_g$ | $\dfrac{\left(2^0 0^0\right)^+ + \left(0^0 2^0\right)^+}{\sqrt{2}}$ |
| $\left(1^1 1^{-1}\right)^+_u$ | $\dfrac{\left(2^0 0^0\right)^+ - \left(0^0 2^0\right)^+}{\sqrt{2}}$ |
| $\left(1^1 1^{-1}\right)^-_u$ | $\left(1^1 1^{-1}\right)^-$ |
| $\left(2^0 0^0\right)^+_g$ | $\left(1^1 1^{-1}\right)^+.$ |

It is now possible, by using this correspondence and our correlation diagram of Fig.7, to obtain the complete correlation between these four *normal* states and the states of a planar configuration for the special symmetric case:

| *linear molecule* | *bent planar* | | *notation in* [10] | $C_{2v}$ *species* |
|---|---|---|---|---|
| $\left(0^0 2^0\right)^+_g$ | $\nu_s$ | (symmetric bend) | $\nu_3$ | $A_1$ |
| $\left(1^1 1^{-1}\right)^+_u$ | $\nu_a$ | (antisymmetric bend) | $\nu_6$ | $B_1$ |
| $\left(1^1 1^{-1}\right)^-_u$ | $\nu_T$ | (torsional fundamental) | $\nu_4$ | $A_2$ |
| $\left(2^0 0^0\right)^+_g$ | $2\nu_T$ | (torsional overtone) | $2\nu_4$ | $A_1$ |

This result is in complete agreement with the receipt of [10]. Moreover, we can generalize this pattern to give explicit formulas for the correlation of any state of arbitrary excitation (not available in [10]). The idea is simply to establish a suitable generalization of formulas (3) and (9) for normal quantum numbers. We therefore introduce:

$$v_\alpha = \frac{(v_4 - |l_4|)}{2}, \quad v_\beta = \frac{(v_5 - |l_5|)}{2}, \quad v_T = |l_4 - l_5| - \frac{(1-p)}{2}. \tag{12}$$

Then, if we interpret $v_\beta$ as the normal $A_1$ bending symmetrical mode, the correlation for the quantum numbers in the bent planar configuration will be given by

$$\begin{aligned} & v_s = v_\beta, \\ & v_T = v_T, \quad v_a = v_\alpha \quad (v_T \text{ odd}), \\ & v_T = 2v_\alpha, \quad v_a = v_T/2 \quad (v_T \text{ even}). \end{aligned} \tag{13}$$

In order to describe other cases of interest, like the symmetric non-planar configuration of the HOOH molecule, the same procedure used in the non-symmetric configuration can be used.

**4. Conclusions**

A simple correlation scheme valid for any kind of four-atomic molecule has been introduced. This receipt can be of great help when trying to parametrize in a continuous way an Hamiltonian operator for the rotovibrating molecule. Preliminary results of this approach have been obtained in the framework of the algebraic vibron model applied to HCNO [14]. In order to show in detail the specific correlation patterns in different practical cases, intermediate correlation laws for free and hindered rotations have been briefly reviewed. As a general comment, it should be emphasized that our method can be extended to larger molecules. For molecules with five or more atoms correlation patterns are extremely complex and in most cases completely unknown. In view of the increasing amount of experimental work in the field of high resolution spectroscopy of medium-sized and large molecules, an unified approach to the problem of the correlation of rotovibrational states can give a non trivial contribution to related theoretical work in this same field.

We want to thank dr.R.Lemus, Prof.B.Winnewisser and especially Prof.F.Iachello for useful discussions during the development of this work.

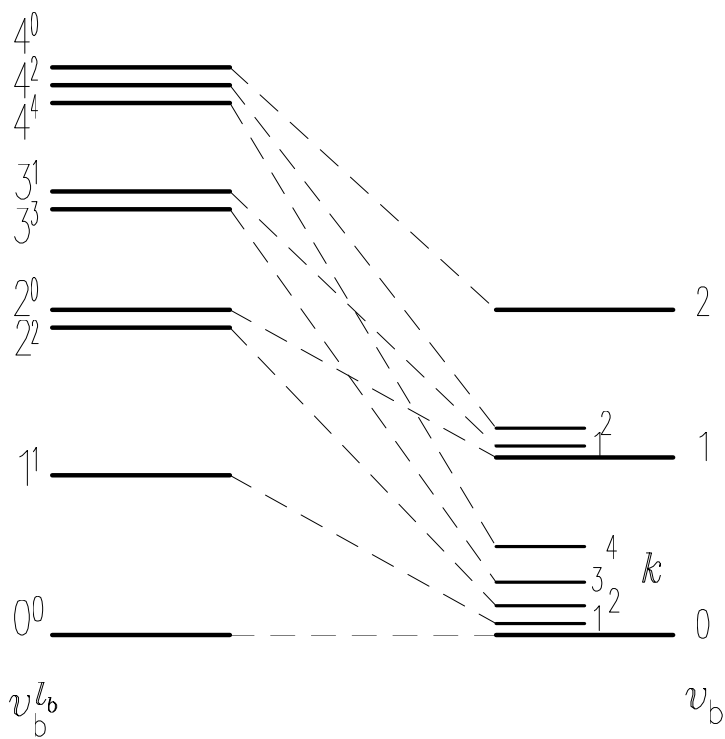

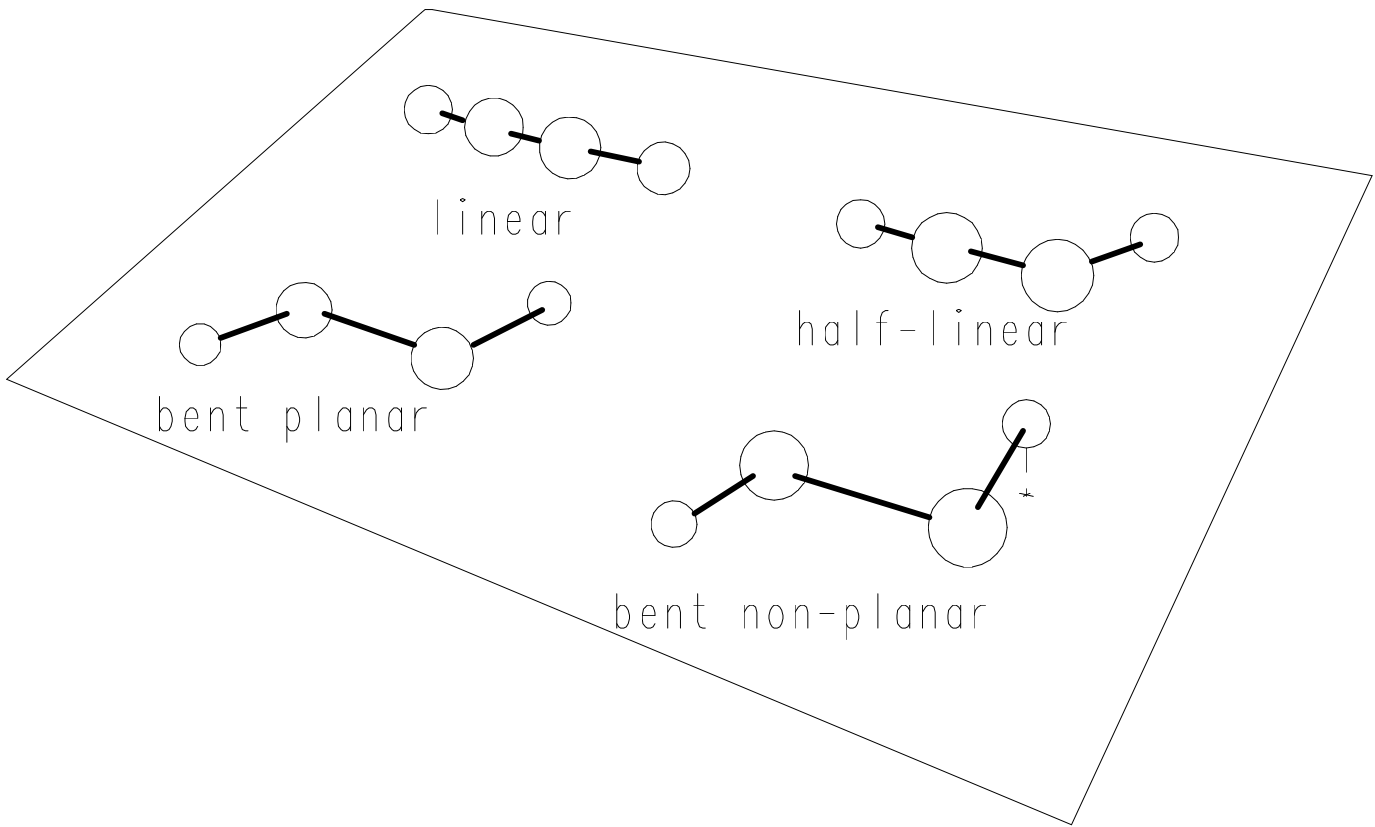

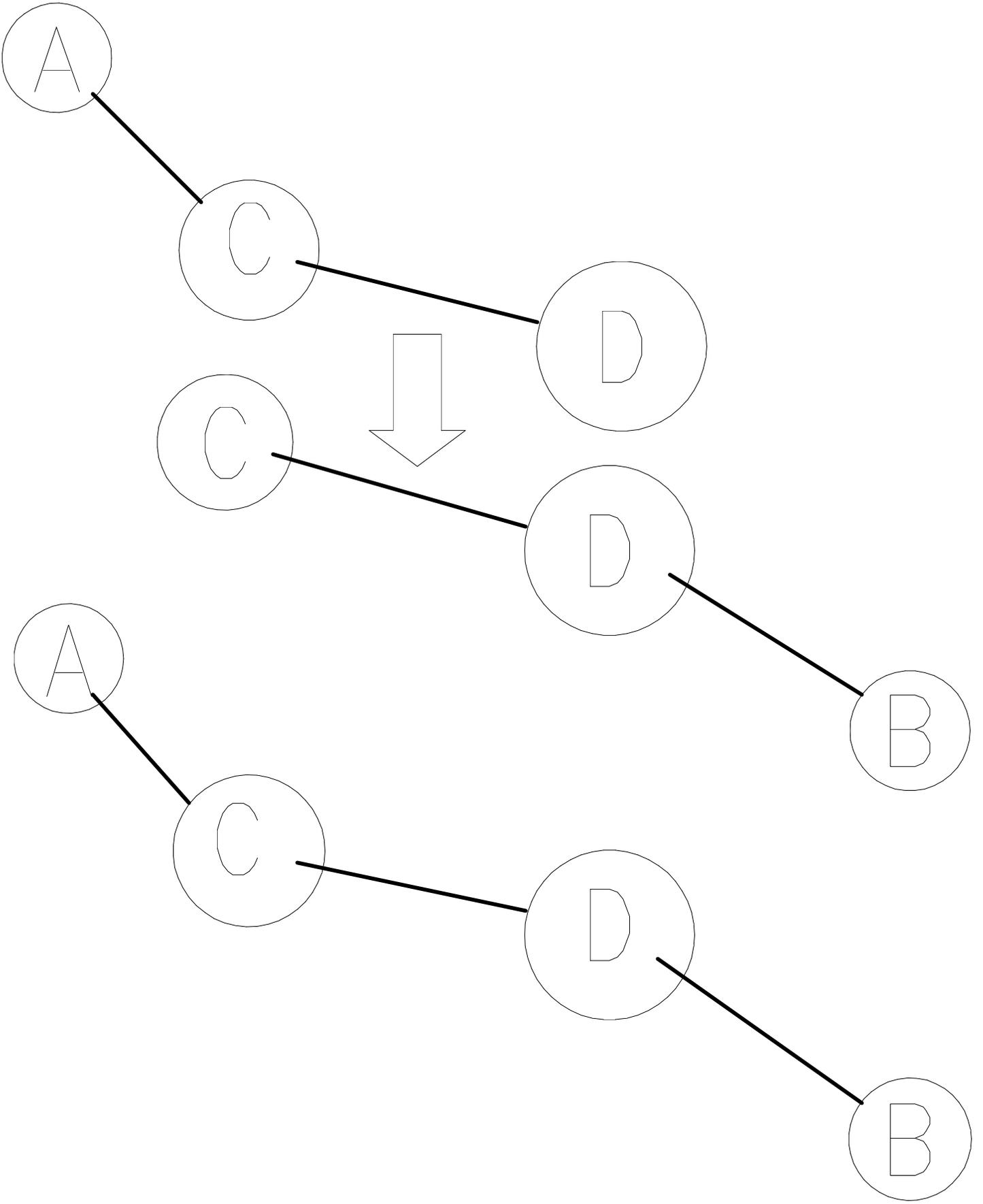

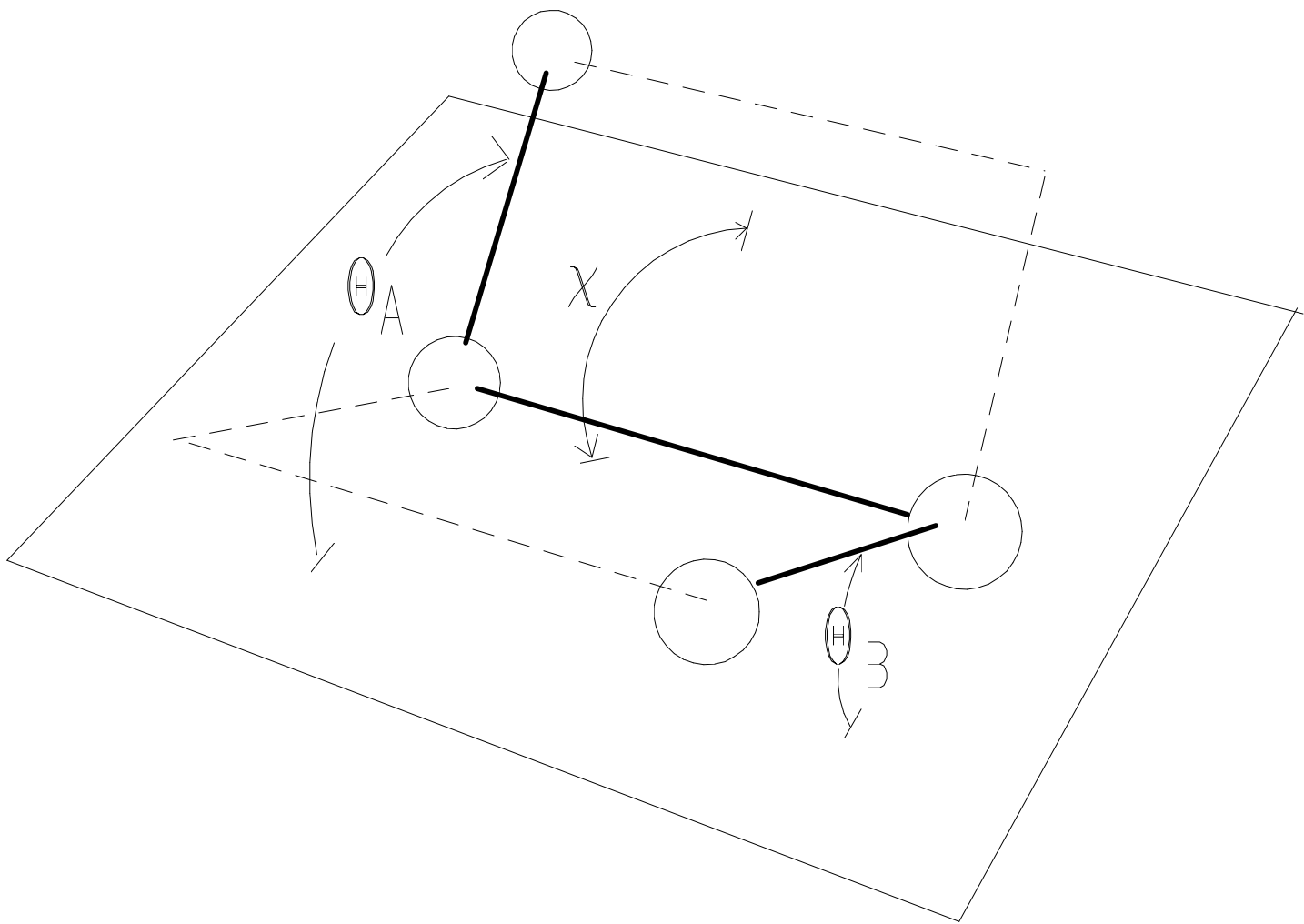

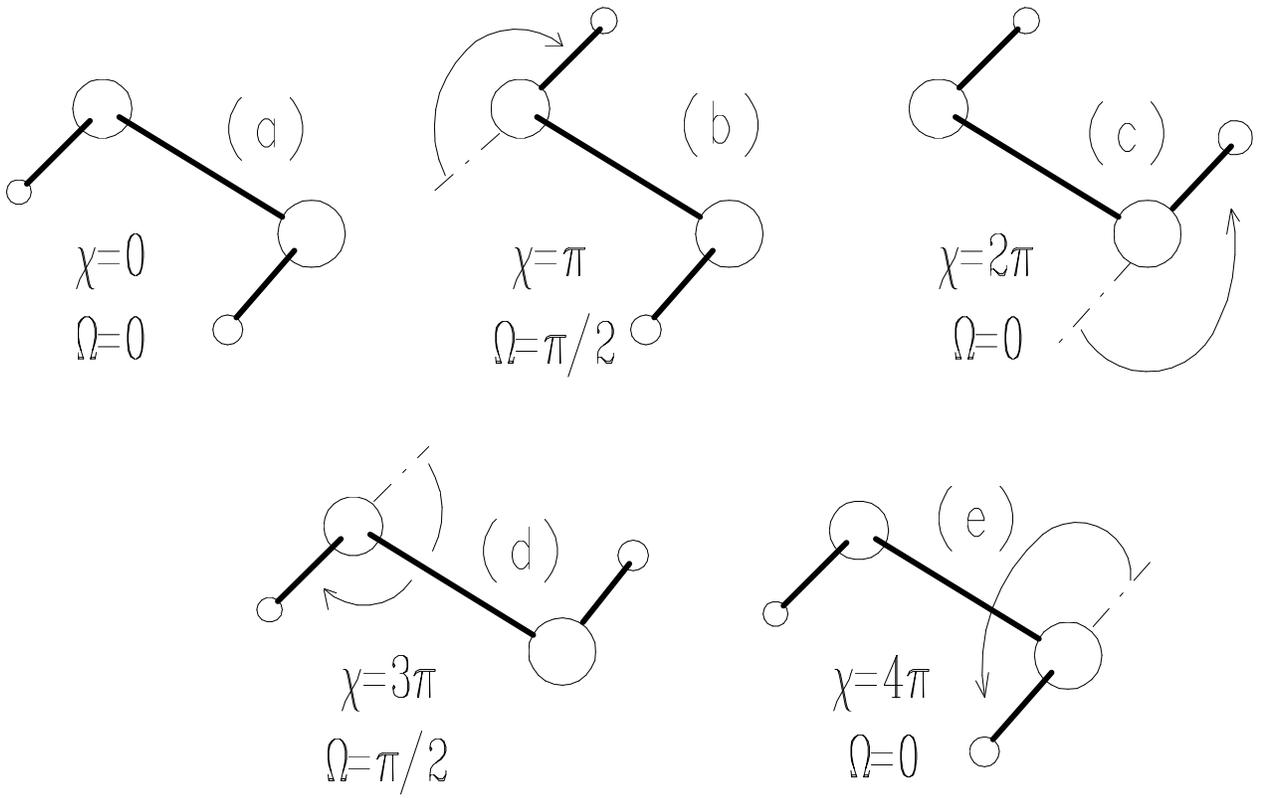

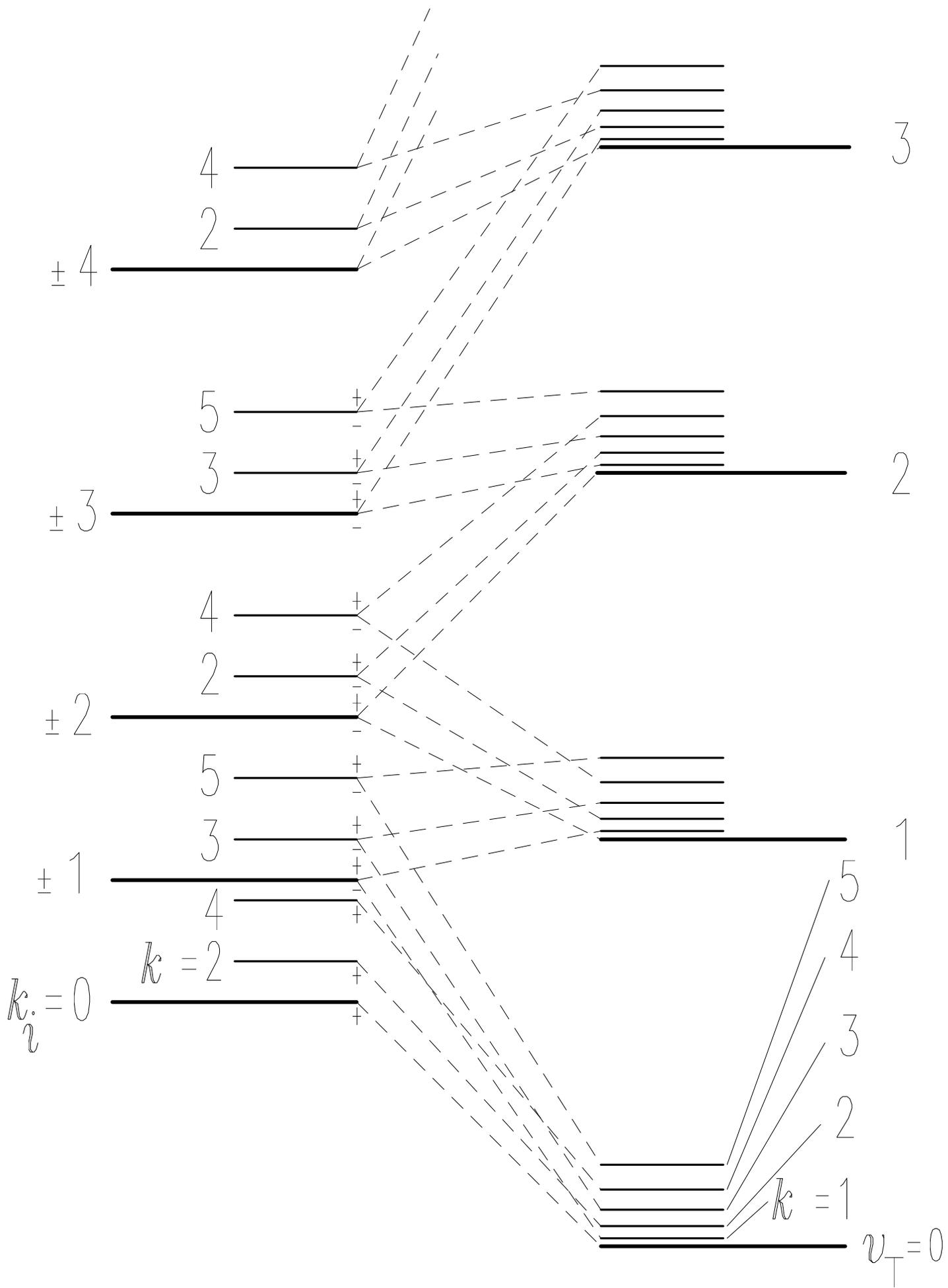

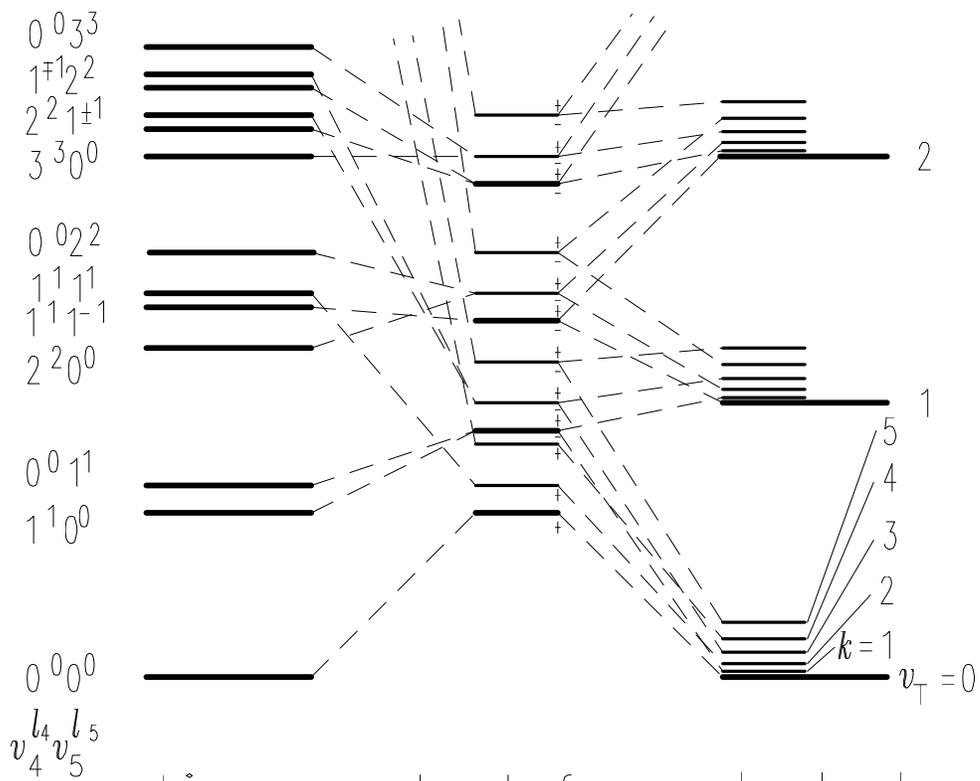

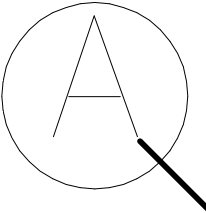
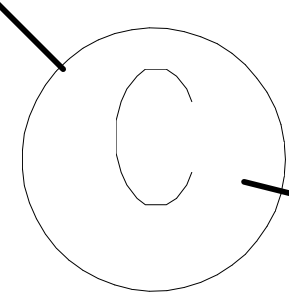
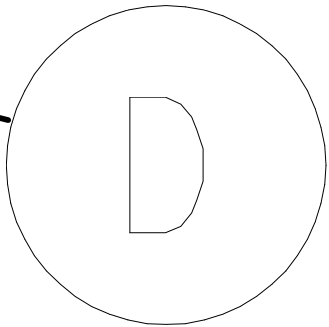
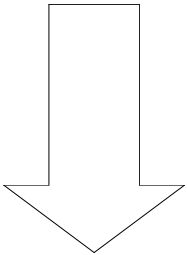
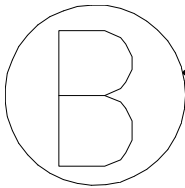
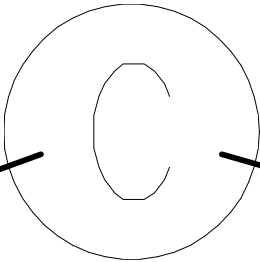
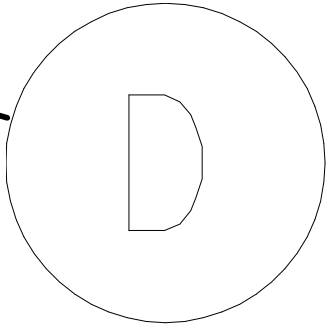
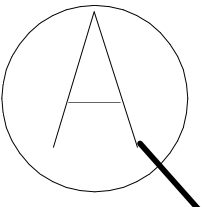
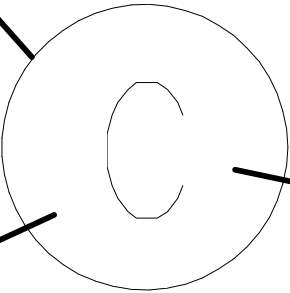
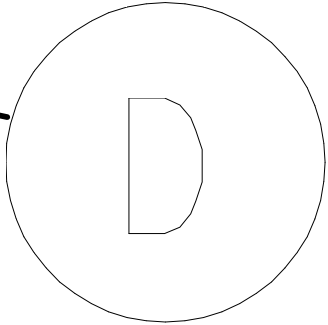
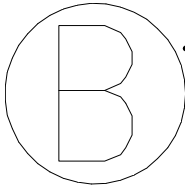

**Figure Captions**

*Figure 1.* Correlation diagram for linear to bent three-atomic molecules [8]. For clearness, only bending states have been included. The axis of energy is not on scale.

*Figure 2.* Different possible geometries of chain four-atomic molecules.

*Figure 3.* Imaginary construction of a chain four-atomic molecule with a pair of three-atomic molecules.

*Figure 4.* Angular coordinates for a bent four-atomic molecule. The reference plane for $\theta_A$, $\theta_B$ is arbitrarily kept fixed in space. The molecule is represented in a generic position, not necessarly an equilibrium position.

*Figure 5.* Internal rotation in a bent four-atomic molecule. The overall periodicity of $4\pi$ for the dihedral angle $\chi$ is explained by comparison of case (c) with case (e), differing by a rotation of the whole molecule through an angle of $\pi$.

*Figure 6.* Correlation diagram, based on Eq.(9), for free to hindered internal rotation in bent four-atomic molecules. For clearness, only few rotational states have been included. The ± label refers to the parity of the molecular state with respect to reflections through the plane defined by the equilibrium position. The torsional quantum number is denoted by $v_T$. The axis of energy is not on scale.

*Figure 7.* Correlation diagram for linear to planar bent four-atomic molecules. The correlation is obtained by considering as an intermediate step the free internal-rotating bent molecule. Labels for these levels (omitted for convenience) are the same as in Fig.6. For clearness, only those levels of the linear molecule related to internal torsional modes have been included. The axis of energy is not on scale.

*Figure 8.* Imaginary construction of a non chain four-atomic molecule with a pair of three-atomic molecules.